\documentclass[preprint,11pt]{elsarticle}

\usepackage{graphicx}
\usepackage{amsmath}
\usepackage{listings}
\usepackage{url}
\usepackage{wasysym}
\usepackage{stfloats}
\usepackage{dcolumn}% Align table columns on decimal point
\usepackage{bm}% bold math
\usepackage{enumitem}
\usepackage{multirow}
\usepackage{relsize}
\usepackage{environ}
\usepackage{placeins}
\usepackage{amssymb}
\usepackage{float}
\usepackage{color}
\usepackage{graphicx}
\usepackage{lipsum}
\usepackage{xspace}
\usepackage{needspace}
\usepackage{hyperref}% add hypertext capabilities

\definecolor{dkgreen}{rgb}{0,0.6,0}
\definecolor{gray}{rgb}{0.5,0.5,0.5}
\definecolor{mauve}{rgb}{0.58,0,0.82}
\newcommand{\dEdx}{d$E$/d$x$\xspace}

\newcommand{\eqnref}[1]{Eq.~\ref{#1}}
\newcommand{\figref}[1]{Fig.~\ref{#1}}

\newcommand{\Secref}[1]{Section~\ref{#1}}
\newcommand{\imagefolder}{./}
\newcommand{\eq}[1]{\begin{align} #1 \end{align}}
\sloppypar
\begin{document}

\begin{frontmatter}


\title{TIdentity module for the reconstruction of the moments of multiplicity distributions}



\author[label1]{Mesut Arslandok}
\author[label1,label2,label3]{Anar Rustamov}

\address[label1]{Physikalisches Institut, University of Heidelberg}
\address[label2]{GSI Helmholtzzentrum f\"{u}r Schwerionenforschung, Darmstadt, Germany}
\address[label3]{National Nuclear Research Center, Baku, Azerbaijan}

\begin{abstract}

In this report a new software module for the reconstruction of the  moments of multiplicity 
distributions of identified particles, the TIdentity module, is presented. The module  
exploits the Identity Method, which allows to circumvent the issues of incomplete particle 
identifications caused by unavoidable overlapping particle identification signals. After 
demonstrating the performance of the module in a number of simulations, we provide a user's 
guide with a detailed description of its functionality. The module can be used in high 
energy nuclear interactions aiming at the determination of the moments of multiplicity 
distributions of identified particles.

\end{abstract}

\end{frontmatter}


%%%%%%%%%%%%%%%%%%%%%%%%%%%%%%%%%%%%%%%%%%%%%%%%%%%%%%%%%%%%%%%%%%%%%%%%%%%%%%%%%%%%
%%%%%%%%%%%%%%%%%%%%%%%%%%%%%%%%%%%%%%%%%%%%%%%%%%%%%%%%%%%%%%%%%%%%%%%%%%%%%%%%%%%%
%%%%%%%%%%%%%%%%%%%%%%%%%%%%%%%%%%%%%%%%%%%%%%%%%%%%%%%%%%%%%%%%%%%%%%%%%%%%%%%%%%%%

\section{\label{sec:intro}Introduction}

The study of the multiplicity distributions of identified particles, in terms of their moments, 
is one of the major topics in high energy nuclear collision experiments. 
For a system in statistical equilibrium, such as strongly interacting matter created in 
relativistic nuclear collisions, specific combinations of the moments of multiplicity 
distributions, referred to as event-by-event fluctuation signals, are directly related to 
the equation of state of strongly interacting matter \cite{Athanasiou:2010kw,Bazavov:2012jq,Shuryak:2000pd}.
The standard approaches of the determination of e-by-e fluctuations signals  are to count the number of  particles in each collision (event). 
However, in  experiments, it is often challenging to identify uniquely the type of every detected particle due to overlapping particle identification signals. For this reason, analyses of identified particle fluctuations are usually performed in a limited kinematic acceptance, where particle identification is relatively reliable, or exploiting additional detectors. These procedures, however, reduce the overall experimental acceptance  and/or particle identification efficiencies, which ultimately pushes the fluctuations of interest to the Poisson limit.
%Therefore, incomplete particle identification is a serious challenge to the precise measurement of 
%fluctuations of the identified-particles. 
\\ \indent
Although it is usually difficult to uniquely identify every detected particle in a given event, to each particle one can still assign a probability of being of a given type. The Identity Method eliminates the effect of incomplete particle identification using these 
probabilities, and allows for the calculation of pure and mixed moments of otherwise unknown particle multiplicity distributions.
% This information will be shown to be sufficient to fully eliminate the effect of incomplete identification.
\\ \indent
The paper is organized as follows. In \Secref{sec:identityMethod} a short introduction to 
the Identity Method and its application to net-particle cumulants are given. 
In \Secref{sec:test} we demonstrate the performance of the TIdentity module in Monte-Carlo 
simulations up to the second order moments. Finally, in \Secref{sec:guide} we discuss the 
implementation of the method in the "TIdentity" module and provide a user's guide.

%%%%%%%%%%%%%%%%%%%%%%%%%%%%%%%%%%%%%%%%%%%%%%%%%%%%%%%%%%%%%%%%%%%%%%%%%%%%%%%%%%%%
%%%%%%%%%%%%%%%%%%%%%%%%%%%%%%%%%%%%%%%%%%%%%%%%%%%%%%%%%%%%%%%%%%%%%%%%%%%%%%%%%%%%
%%%%%%%%%%%%%%%%%%%%%%%%%%%%%%%%%%%%%%%%%%%%%%%%%%%%%%%%%%%%%%%%%%%%%%%%%%%%%%%%%%%%

\section{\label{sec:identityMethod}The Identity Method}

The Identity Method was proposed in Ref.~\cite{Gazdzicki:2011xz} as a solution to the aforementioned misidentification problem 
for the analysis of events with two different particle species. The method was extended  
to calculate the second moments of the multiplicity distributions of more than two particle species in Ref.~\cite{Gorenstein:2011hr}.
Finally, in Ref.~\cite{Rustamov:2012bx}, it was generalized to the first and higher moments of  multiplicity distributions for an 
arbitrary number of particle species and further reexamined in Refs.~\cite{Pruneau:2017fim,Pruneau:2018glv}. The first experimental results using the Identity Method were published by the NA49 \cite{Anticic:2013htn} and ALICE \cite{Acharya:2017cpf,Rustamov:2017lio} collaborations at CERN. 
\\ \indent
Instead of identifying every detected particle event-by-event, the Identity Method calculates the moments of particle 
multiplicity distributions by means of an unfolding procedure using only two basic experimentally measurable track-by-track and event-by-event 
quantities, $\omega$ and $W$, respectively. They are defined as 
\\
%%%%%%%%%%%%%%%%%%%%%%%%%%%%%%%%%%%%%%%%%%%%%%%%%%%%%%%%%%%%%%%%%%%%%%%%%%%%%%%%%%%%
\begin{align} 
  \omega_{j}(x_{i}) & = \dfrac{\rho_{j}(x_{i})}{\rho(x_{i})} \in[0,1],  \quad  \rho(x_{i}) = \sum_{j}\rho_{j}(x_{i}), \label{eq:omegaRho} \\ 
  \quad W_{j}  & \equiv \sum_{i=1}^{N(n)} \omega_{j}(x_{i}) \label{eq:Ws}.
\end{align}
\\
%%%%%%%%%%%%%%%%%%%%%%%%%%%%%%%%%%%%%%%%%%%%%%%%%%%%%%%%%%%%%%%%%%%%%%%%%%%%%%%%%%%%
\noindent where $x_{i}$ stands for the \dEdx of a given track $i$, $\rho_{j}(x)$ is the \dEdx 
distribution of particle type $j$ within a given phase-space bin and $N(n)$ is the number of tracks in 
the $n^{\rm th}$ event. We note that, $x$ can refer to any experimentally measured particle property related to its identity. The $\omega_{j}$ quantity is the probability for a given track to be of particle type $j$ and $W_{j}$ represents a proxy for its multiplicity in a given event. 
Thus, in case of a perfect identification one gets $W_{j} = N_{j}$ with $N_{j}$ denoting the number of particle type $j$ in a given event. 
\\ \indent
As the $W_{j}$ quantities are calculated for each event, by straightforward averaging over all events the moments of $W_{j}$ distributions, including mixed moments between different particle types can be easily reconstructed. The Identity Method provides a set of linear equations, which relates the moments of the $W_{j}$ distributions to the looked for moments of multiplicity distributions~\cite{Rustamov:2012bx}. 
%
%%%%%%%%%%%%%%%%%%%%%%%%%%%%%%%%%%%%%%%%%%%%%%%%%%%%%%%%%%%%%%%%%%%%%%%%%%%%%%%%%%%%
%%%%%%%%%%%%%%%%%%%%%%%%%%%%%%%%%%%%%%%%%%%%%%%%%%%%%%%%%%%%%%%%%%%%%%%%%%%%%%%%%%%%
%%%%%%%%%%%%%%%%%%%%%%%%%%%%%%%%%%%%%%%%%%%%%%%%%%%%%%%%%%%%%%%%%%%%%%%%%%%%%%%%%%%%
%
\section {The TIdentity Module}
The TIdentity module is developed for the calculation of the moments of particle multiplicity distributions for an arbitrary 
number of particle species using the Identity Method described in the previous section. The code works within the ROOT framework~\cite{Brun:1997pa} (versions 5 and 6) and is accessible online via:
\begin{lstlisting}[mathescape]
git clone https://github.com/marslandALICE/TIdentityModule.git
\end{lstlisting}
The TIdentity module requires two inputs: (i) the track-by-track measurements in each event, such as $x_{i}$ quantities in Eq.~\ref{eq:omegaRho}; (ii) the distribution of the  identification signal for  each particle type, like $\rho_{j}(x)$ of Eq.~\ref{eq:omegaRho}.
The current version of the module contains calculations of all pure and mixed moments of particle multiplicity distributions up to the second order. 
To calculate the $n^{th}$ moment of  multiplicity distributions, the TIdentity module uses all pure and mixed $(n-1)^{th}$ moments and 
$\rho_{j}(x)$ functions. The details on practical usage of the TIdentity module are given in the user's guide, provided in section~\ref{sec:guide}.

%Further mathematical insights of the method and its verification with Monte-Carlo simulations can be found in Refs.~\cite{Gorenstein:2011hr, Rustamov:2012bx}. 

\subsection{\label{sec:test} Test on simulated data}
\begin{figure}[H]
  \centering
  \includegraphics[width=12.5cm]{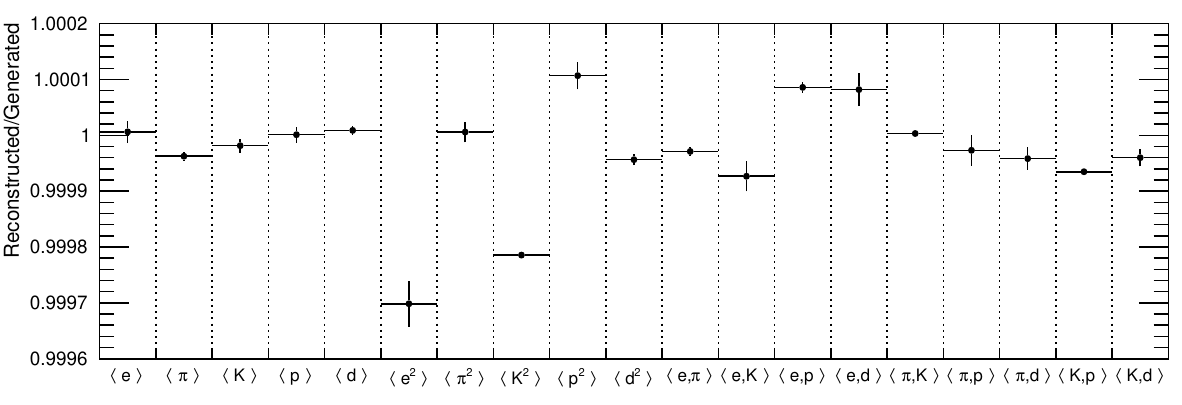}
  \caption{Ratio of the reconstructed moments to the generated ones obtained with the TIdentity module.}
  \label{fig:genVSrec}
\end{figure}

 Here we demonstrate the usage of the TIdendity module on a simulated data sample of 10 million events. We assume that particle 
identification is achieved by measuring the specific energy loss \dEdx. In each event we generate five different particle species; electron (e), pion ($\pi$), proton (p), kaon (K) and deuteron (d), as well as their anti-particles. The multiplicity distributions of these particles are 
simulated according to the Poisson distribution with different mean values. 
\\ \indent 
%%%%%%%%%%%%%%%%%%%%%%%%%%%%%%%%%%%%%%%%%%%%%%%%%%%%%%%%%%%%%%%%%%%%%%%%%%%%%%%%%%%%
\begin{figure}
  \centering
  \includegraphics[width=12.8cm]{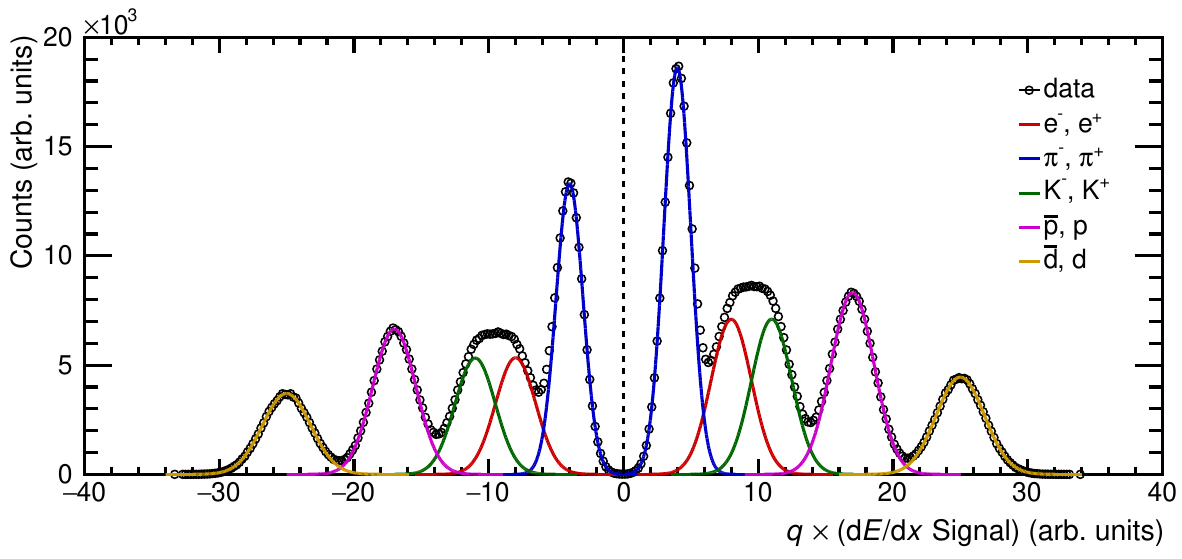}
  \caption{(Color online) A Monte-Carlo simulation of particle energy losses for electrons, pions, kaons, protons and deuterons. Anti-particle 
  and particle distributions are shown in the left and right panels, respectively.}
  \label{fig:dEdxDist}
\end{figure}
%%%%%%%%%%%%%%%%%%%%%%%%%%%%%%%%%%%%%%%%%%%%%%%%%%%%%%%%%%%%%%%%%%%%%%%%%%%%%%%%%%%%
The input information to the simulation is a \dEdx distribution for each particle type
(see \figref{fig:dEdxDist}) and mean multiplicities of particles (anti-particles), which are taken to be 8 (6), 14 (10), 8 (6), 10 (8) and 6 (5) for electrons, pions, kaons, protons and deuterons, respectively. The mean multiplicity of kaons (anti-kaons) is taken to be the same as for electrons (positrons) in each event in order to demonstrate that the method functions also for correlated particle production.
\\ \indent
The simulation process comprises the following steps: (i) in each event we 
randomly generate particle multiplicities from the Poisson distribution using their mean values; (ii) we generate a \dEdx value for each particle in an event from the inclusive distribution $\rho_j(x)$ with Gaussian shape for the corresponding particle type $j$. Using this simulated data we reconstruct all the first and second moments of the multiplicity distributions with the TIdentity module. 
\\ \indent
%%%%%%%%%%%%%%%%%%%%%%%%%%%%%%%%%%%%%%%%%%%%%%%%%%%%%%%%%%%%%%%%%%%%%%%%%%%%%%%%%%%%
\begin{figure}[H]
  \centering
  \includegraphics[width=11cm]{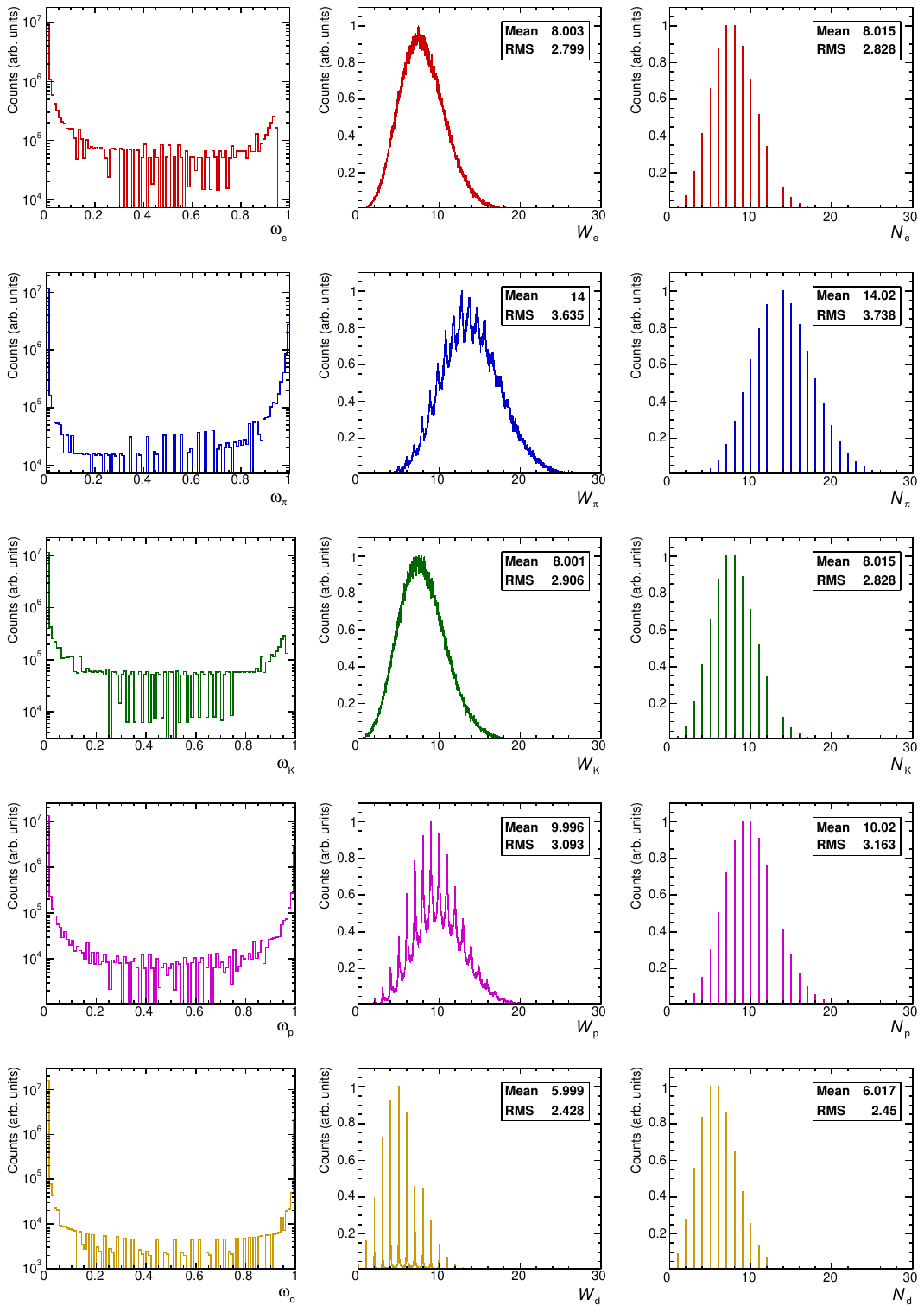}
  \caption{(Color online) The distribution of track variables $\omega$ (first column), event variables $W$ (second column) and the true 
  multiplicity distributions $N$ (third column) of electrons, pions, kaons, protons and deuterons.}
  \label{fig:WsNs}
\end{figure}
%%%%%%%%%%%%%%%%%%%%%%%%%%%%%%%%%%%%%%%%%%%%%%%%%%%%%%%%%%%%%%%%%%%%%%%%%%%%%%%%%%%%
For each simulated particle in an event we calculate the identity variables $\omega_{j}(x)$ according to \eqnref{eq:omegaRho}, where $j$ stands for the particle type. Next, for each particle type, we determine the event variable $W_{j}$ (cf.~\eqnref{eq:Ws}). The left panel of \figref{fig:WsNs} shows the distributions of the identity variables $\omega_{j}$ for different particle species. In the middle and right columns of \figref{fig:WsNs} the distributions of the $W_{j}$ quantities and the true multiplicity distributions $N_{j}$ are plotted, respectively. 
\\ \indent
The mean values of $W_{j}$ and $N_{j}$ distributions are the same, which is reflected in the middle and right panels of Fig.~\ref{fig:WsNs}.
Moreover, the $W$ distributions of electrons and kaons have a smooth behavior, since their $W$ values are dominated by the overlapping regions of the \dEdx distributions, thus they predominantly take non-integer values. The distributions of pions, protons and deuterons exhibit evident structures because the \dEdx distributions of these particles are relatively well separated from those of electrons and kaons as illustrated in the middle panel of \figref{fig:WsNs}. These structures are nothing  but the smeared original multiplicity distributions because of the non-ideal particle identification, i.e., because of the overlapping \dEdx distributions of different particle species. 
\\ \indent 
%%%%%%%%%%%%%%%%%%%%%%%%%%%%%%%%%%%%%%%%%%%%%%%%%%%%%%%%%%%%%%%%%%%%%%%%%%%%%%%%%%%%
\begin{figure}[H]
  \centering
  \includegraphics[width=8cm]{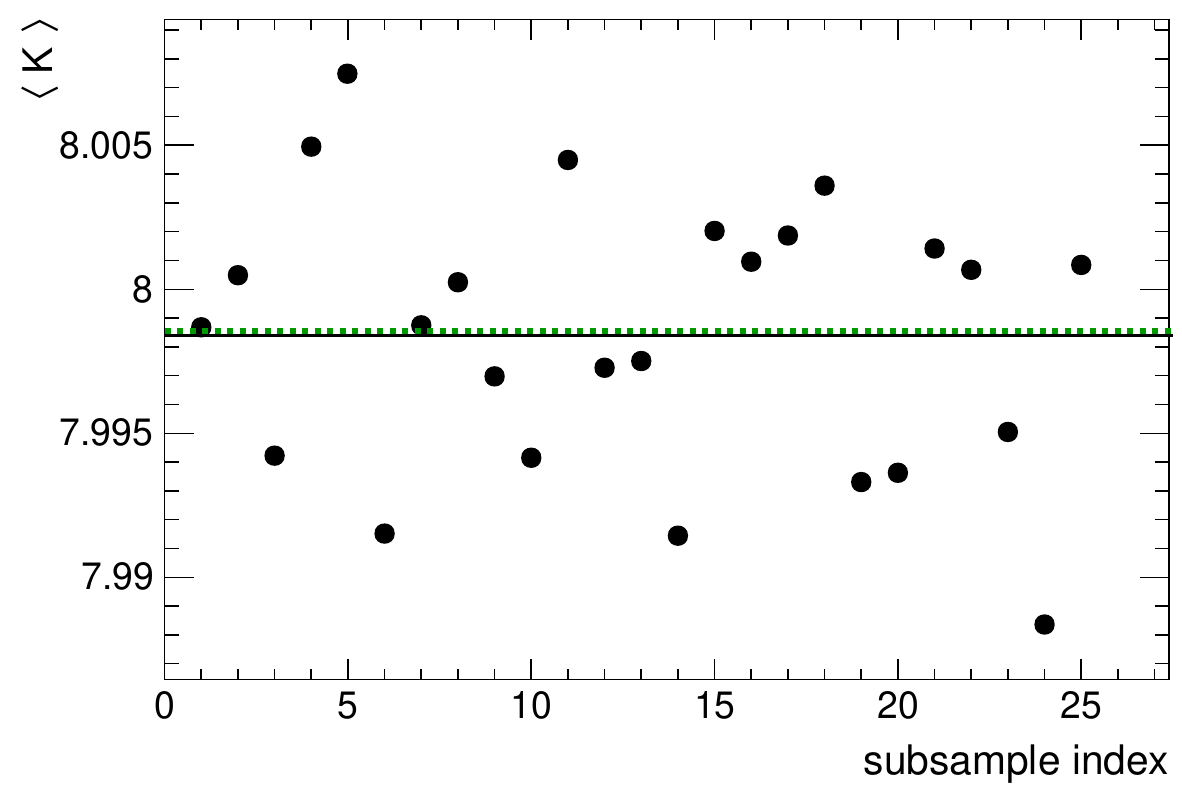}
  \caption{Reconstructed values of $\langle K \rangle$ for 25 subsamples. The dashed green lines indicate the corresponding averaged values of $\langle K \rangle$ over subsamples and the black solid line indicates the generated value of $\langle K \rangle$.}
  \label{fig:subsample}
\end{figure}
%%%%%%%%%%%%%%%%%%%%%%%%%%%%%%%%%%%%%%%%%%%%%%%%%%%%%%%%%%%%%%%%%%%%%%%%%%%%%%%%%%%%
Using Eq.~16 of Ref~\cite{Rustamov:2012bx}, we reconstruct 
all first moments of the particles from the corresponding moments of the $W$ quantities. The statistical uncertainties of the moments are evaluated with the subsample method. We first divided the data set into $n=25$ random subsamples, and reconstructed the moments for each subsample, as 
illustrated in \figref{fig:subsample}. Next, we calculated the statistical error for a given moment $M$ with the following formula:
%%%%%%%%%%%%%%%%%%%%%%%%%%%%%%%%%%%%%%%%%%%%%%%%%%%%%%%%%%%%%%%%%%%%%%%%%%%%%%%%%%%%
\begin{equation}
  \sigma_{\langle M \rangle} = \dfrac{\sigma}{\sqrt{n}}
\end{equation}
\\
where 
\\
\begin{equation}
  \langle M \rangle = \dfrac{1}{n}\sum M_{i}, \quad 
  \sigma = \sqrt{\dfrac{\sum (M_{i} - \langle M \rangle )^2}{n-1}}.
\end{equation}
%%%%%%%%%%%%%%%%%%%%%%%%%%%%%%%%%%%%%%%%%%%%%%%%%%%%%%%%%%%%%%%%%%%%%%%%%%%%%%%%%%%%
\\ \indent
Finally, as shown in \figref{fig:genVSrec}, we compare the reconstructed moments calculated with the TIdentity module to the generated ones. The precision of the agreement is better than 0.1\permil. We note that, the observed tiny deviations from unity are due to the systematic uncertainties stemming from the finite bin width of the \dEdx distributions and the precision of the $\rho_{j}$ fit functions. 
%%%%%%%%%%%%%%%%%%%%%%%%%%%%%%%%%%%%%%%%%%%%%%%%%%%%%%%%%%%%%%%%%%%%%%%%%%%%%%%%%%%%
%%%%%%%%%%%%%%%%%%%%%%%%%%%%%%%%%%%%%%%%%%%%%%%%%%%%%%%%%%%%%%%%%%%%%%%%%%%%%%%%%%%%

%%%%%%%%%%%%%%%%%%%%%%%%%%%%%%%%%%%%%%%%%%%%%%%%%%%%%%%%%%%%%%%%%%%%%%%%%%%%%%%%%%%%
%%%%%%%%%%%%%%%%%%%%%%%%%%%%%%%%%%%%%%%%%%%%%%%%%%%%%%%%%%%%%%%%%%%%%%%%%%%%%%%%%%%%
%%%%%%%%%%%%%%%%%%%%%%%%%%%%%%%%%%%%%%%%%%%%%%%%%%%%%%%%%%%%%%%%%%%%%%%%%%%%%%%%%%%%

\subsection{\label{sec:netpart} Net-particle cumulants}
Net-particle cumulants are in particular important to address fluctuations of conserved charges~\cite{Bazavov:2012jq}. 
In this section we demonstrate a procedure for reconstructing net-particle cumulants of the second order. The higher order cumulants of net-particles 
can be reconstructed in a similar way. Given the probability distribution $P(n_{j})$ of particle type $j$, its $r^{th}$ raw moment is defined as:
\eq{
\label{raw_moments}
&\left<n_{j}^{r}\right>= \sum_{n_{j}=0}^{\infty}\ n_{j}^{r}P(n_{j}).
}
 The second cumulant of net-particles $\Delta n_{j} = n_{j} - n_{\bar{j}}$, with $n_{j}$ and $n_{\bar{j}}$ referring to the number of particles and anti-particles, reads:
\begin{equation} \label{eq:kumulants_definition}
 \kappa_{2} (\Delta n_{j}) = \langle\Delta n_{j}^2\rangle - \langle\Delta n_{j}\rangle^{2}, 
\end{equation}
% \eq{
% \label{kumulants_definition}
% %  &\kappa_{1}(\Delta n_{B}) = \sum_{\Delta n_{B}=0}^{\infty}\Delta n_{B}P(\Delta n_{B})=\left<\Delta n_{B}\right> , \\
% &\kappa_{2} (\Delta n) = \left<\Delta n^2\right> - \left<\Delta n\right>^{2}
% }
which can be represented  as a sum of corresponding cumulants for
particles plus the correlation term for the joint probability distributions of particles and antiparticles $P(n_{j}, n_{\bar{j}})$
\begin{equation}
\kappa_{2}(\Delta n_{j})= \kappa_{2}(n_{j})+\kappa_{2}(n_{\bar{j}}) - 2\left(\langle n_{j}n_{\bar{j}} \rangle -
\langle n_{j}\rangle \langle n_{\bar{j}} \rangle\right),
\label{sec_cum_2}
\end{equation}
where the mixed cumulant $\left<n_{j}n_{\bar{j}}\right>$ is defined as
\eq{
\label{mixed_cumulant}
\langle n_{j}n_{\bar{j}}\rangle = \sum_{n_{j}=0}^{\infty}\sum_{n_{\bar{j}}=0}^{\infty}n_{j}n_{\bar{j}}P(n_{j},n_{\bar{j}}).
}
\indent Here, we discuss two distinct ways of calculating this correlation term. In the first case, we run the program for 3 different settings: only for particles, only for antiparticles and for the sum of particles and antiparticles. The latter means that the probabilities calculated according to \eqnref{eq:omegaRho} 
are summed up for particles and antiparticles. Hence, using  the reconstructed  second order moments, $\langle n_{j}^{2} \rangle$, $\langle n_{\bar{j}}^{2} \rangle$ and $\big \langle \left(n_{j}+n_{\bar{j}}\right)^2 \big \rangle$, the correlation term can be easily calculated: 
\eq{
\label{mixed_cumulant2}
\langle n_{j}n_{\bar{j}} \rangle = \frac{1}{2} \left( \big \langle \left( n_{j}+n_{\bar{j}} \right)^2 \big \rangle - \langle n_{j}^{2} \rangle - \langle n_{\bar{j}}^{2} \rangle \right).
}
Alternatively one can directly calculate the second moment of the $n_{j}-n_{\bar{j}}$ distribution entering \eqnref{eq:kumulants_definition}: 
\eq{
\label{mixed_cumulant3}
\langle \Delta n_{j}^{2}\rangle = 2\langle n_{j}^{2} \rangle + 2\langle n_{\bar{j}}^2 \rangle - \big \langle \left(n_{j}+n_{\bar{j}}\right)^2 \big \rangle.
}

We note that this approach was used in the ALICE net-proton analysis~\cite{Rustamov:2017lio}. In the second case, the correlation term $\left<n_{j}n_{\bar{j}}\right>$ can be obtained directly by using the particles and antiparticles simultaneously.  
This means that the probabilities defined in \eqnref{eq:omegaRho} are calculated separately for particles and antiparticles. Consequently this yields the moments of both  $W_{j}$ and $W_{\bar{j}}$ distributions including the mixed moments of their joint distributions. This allows for the determination of the correlation term $\langle n_{j}n_{\bar{j}} \rangle$ mentioned above. We note that both of these methods should yield identical results. However, their performance may depend on the number of analyzed particles, because in the second case the effective number of particles doubles, which in turn may lead to computational difficulties.

%%%%%%%%%%%%%%%%%%%%%%%%%%%%%%%%%%%%%%%%%%%%%%%%%%%%%%%%%%%%%%%%%%%%%%%%%%%%%%%%%%%%
%%%%%%%%%%%%%%%%%%%%%%%%%%%%%%%%%%%%%%%%%%%%%%%%%%%%%%%%%%%%%%%%%%%%%%%%%%%%%%%%%%%%
%%%%%%%%%%%%%%%%%%%%%%%%%%%%%%%%%%%%%%%%%%%%%%%%%%%%%%%%%%%%%%%%%%%%%%%%%%%%%%%%%%%%

\section{\label{sec:guide}User's Guide}
%The TIdentity module is developed for the calculation of the moments of particle multiplicity distributions for an arbitrary 
%number of particle species using the Identity Method. The code works within the ROOT framework~\cite{Brun:1997pa} (versions 5 and 6) and is accessible online via:
%\begin{lstlisting}[mathescape]
%git clone https://github.com/marslandALICE/TIdentityModule.git
%\end{lstlisting}
%It requires two input files: 
%the first one contains a TTree object with the track-by-track information of each event, while the second one is a look-up table of the 
%line shapes, i.e.\ the fit functions of the \dEdx distributions of each particle species. To demonstrate the usage %of the code two example macros are 
%provided: (i) \textit{testIdentity\_Sign.C} calculates the moments for particles and antiparticles separately or together, 
%(ii) \textit{testIdentity\_NetParticles.C} calculates the moments of particles and anti-particles simultaneously. 
%Further details are given in a README file. 

%The TIdentity module is  is accessible online via:

%\begin{lstlisting}[mathescape]composed of three main classes:

%%%%%%%%%%%%%%%%%%%%%%%%%%%%%%%%%%%%%%%%%%%%%%%%%%%%%%%%%%%%%%%%%%%%%%%%%%%%%%%%%%%%
%%%%%%%%%%%%%%%%%%%%%%%%%%%%%%%%%%%%%%%%%%%%%%%%%%%%%%%%%%%%%%%%%%%%%%%%%%%%%%%%%%%%
%%%%%%%%%%%%%%%%%%%%%%%%%%%%%%%%%%%%%%%%%%%%%%%%%%%%%%%%%%%%%%%%%%%%%%%%%%%%%%%%%%%%

%\subsection{Running the Code}
The current version of the TIdentity module is composed of the following classes:
\begin{itemize}
  \item \textbf{TIdentityBase} is the base class, from which the other classes inherit.
  \item \textbf{TIdentityFunctions} is used to process the line-shapes.
  \item \textbf{TIdentity2D} is the main steering class for the calculation of the second order moments.
\end{itemize}
An instance of the TIdentity2D  class can be created via its constructor:
\begin{lstlisting}[mathescape]
TIdentity2D::InitIden2D(Int_t size)       
\end{lstlisting}
where ``size" is the number of particle species used in a given analysis. Below, important 
functions that are involved in the TIdentity2D 
class are introduced in detail. 

%%%%%%%%%%%%%%%%%%%%%%%%%%%%%%%%%%%%%%%%%%%%%%%%%%%%%%%%%%%%%%%%%%%%%%%%%%%%%%%%%%%%
%%%%%%%%%%%%%%%%%%%%%%%%%%%%%%%%%%%%%%%%%%%%%%%%%%%%%%%%%%%%%%%%%%%%%%%%%%%%%%%%%%%%
%%%%%%%%%%%%%%%%%%%%%%%%%%%%%%%%%%%%%%%%%%%%%%%%%%%%%%%%%%%%%%%%%%%%%%%%%%%%%%%%%%%%

The TIdentity module works within the ROOT framework~\cite{Brun:1997pa} (versions 5 and 6) and is accessible online via:

\begin{lstlisting}[mathescape]
git clone https://github.com/marslandALICE/TIdentityModule.git
\end{lstlisting}

In the first step, the  the module  has to be compiled  as it is specified in the provided README file. 
The following input files are needed: 
(i) a file containing a TTree object with the track-by-track information of each event;
(ii) functional form of the \dEdx distributions of each particle species. 

Next, the following three functions in the test macros need to be modified according to the needs of a given analysis:
\begin{itemize}
  \item \textbf{\texttt{InitializeObjects()}} is used to initialize an output tree that contains the moments and other relevant information, as well as the pointer arrays, which keep the line-shapes in the memory.
  \item \textbf{\texttt{ReadFitParamsFromLineShapes(TString)}} is used to read the input lookup table that contains the line-shapes. The line-shapes can 
  be stored in either TF1 or TH1D objects, where the latter is advantageous in terms of CPU time but requires an optimum selection of the histogram binning. 
  %Therefore, the user needs to tune the number of bins in the TH1D objects to achieve the desired precision.
  \item \textbf{\texttt{EvalFitValue(Int{\_}t particle, Double{\_}t x)}} is used to calculate $\rho_{j}(x)$ values for each track. The pointer to this function is set by the ``SetFunctionPointers" function.
\end{itemize}
The calculation of the moments is carried out by means of the following public member functions of the TIdentity2D class:
\begin{itemize}
  \item \textbf{\texttt{TIdentity2D::SetFileName(TString $dataTreeFilePath$)}} sets the path to the input data tree that contains track-by-track information.
  \item \textbf{\texttt{TIdentity2D::GetTree(Long{\_}t \&$n$, TString)}} reads the input tree-branches and provides access to them via the corresponding ``Get" functions.
  \item \textbf{\texttt{TIdentity2D::SetBranchNames(const Int{\_}t $tmpNBranches$, TString $tmpBranchNameArr$[])}} is used to introduce the data tree structure 
  to the TIdentity code.
  \item \textbf{\texttt{TIdentity2D::SetFunctionPointers(fptr)}} sets a pointer to the ``EvalFitValue" function.
  \item \textbf{\texttt{TIdentity2D::GetEntry(Int{\_}t $entry$)}} retrieves the entries of the data tree.
  \item \textbf{\texttt{TIdentity2D::GetBins(const Int{\_}t $nExtraBins$, Double{\_}t *)}} 
allows access to the branches of the input tree. 
  \item \textbf{\texttt{TIdentity2D::Reset()}} resets all counters to zero.
  \item \textbf{\texttt{TIdentity2D::SetUseSign(Int{\_}t $sign$)}} sets the particle type to be analysed; particle($sign=1$) or anti-particle($sign=-1$). One 
  can also run the code for the sum of particles and anti-particles by setting $sign=0$.
  \item \textbf{\texttt{TIdentity2D::AddParticles()}} accumulates the $\omega$ quantities which are used to calculate $W$ quantities.
  \item \textbf{\texttt{TIdentity2D::Finalize()}} calculates the moments of the $W$ quantities.
  \item \textbf{\texttt{TIdentity2D::AddIntegrals(Int{\_}t $sign$)}} calculates the elements of the Identity matrix.   
  \item \textbf{\texttt{TIdentity2D::CalcMoments()}} calculates all moments up to the second order for each particle type.
  \item \textbf{\texttt{TIdentity2D::GetMean(Int{\_}t $j$)}} is used to retrieve the first moment of particle type $j$; $\langle N_{j} \rangle$.
  \item \textbf{\texttt{TIdentity2D::GetSecondMoment(Int{\_}t $j$)}} is used to retrieve the second moment of particle type $j$; $\langle N_{j}^{2} \rangle$
  \item \textbf{\texttt{TIdentity2D::GetMixedMoment(Int{\_}t $i$, Int{\_}t $j$)}} is used to retrieve the mixed moment of particle types $i$ and $j$; 
  $\langle N_{i}N_{j} \rangle$
\end{itemize}

For further details two example macros are 
provided in the module: (i) \textit{testIdentity\_Sign.C} calculates the moments for particles and antiparticles separately or together, 
(ii) \textit{testIdentity\_NetParticles.C} calculates the moments of particles and anti-particles simultaneously. 
Further details are given in a README file. 

%%%%%%%%%%%%%%%%%%%%%%%%%%%%%%%%%%%%%%%%%%%%%%%%%%%%%%%%%%%%%%%%%%%%%%%%%%%%%%%%%%%%
%%%%%%%%%%%%%%%%%%%%%%%%%%%%%%%%%%%%%%%%%%%%%%%%%%%%%%%%%%%%%%%%%%%%%%%%%%%%%%%%%%%%
%%%%%%%%%%%%%%%%%%%%%%%%%%%%%%%%%%%%%%%%%%%%%%%%%%%%%%%%%%%%%%%%%%%%%%%%%%%%%%%%%%%%

\section{\label{sec:summary}Conclusions}
In summary, we developed  the TIdentity module to reconstruct  the moments of particle multiplicity distributions. 
%To demonstrate the performance of the method, 
The method was tested in Monte-Carlo simulations with ten million events containing five 
particle (anti-particle) species with different multiplicities. In order to account for 
particle misidentification, we allowed for overlaps of the individual particle energy 
loss distributions. The obtained moments are consistent with the generated ones with a 
precision better than 0.1\permil.
\\ \indent 
Even though the Identity Method was developed specifically for the analysis of event-by-event fluctuations of particle multiplicities, in particular using \dEdx as the identity variable, it can be used for different studies such as reconstruction of particle spectra. Moreover, instead of \dEdx, other observables e.g.\ invariant mass or time-of-flight measurements can be used as the identity variable. 
\\ \indent
The current version of the code unfolds the moments up to the second order, which will be extended to  higher-order moments in the future releases. We further note that, the method can also be exploited in different fields facing similar problems.

%%%%%%%%%%%%%%%%%%%%%%%%%%%%%%%%%%%%%%%%%%%%%%%%%%%%%%%%%%%%%%%%%%%%%%%%%%%%%%%%%%%%
%%%%%%%%%%%%%%%%%%%%%%%%%%%%%%%%%%%%%%%%%%%%%%%%%%%%%%%%%%%%%%%%%%%%%%%%%%%%%%%%%%%%
%%%%%%%%%%%%%%%%%%%%%%%%%%%%%%%%%%%%%%%%%%%%%%%%%%%%%%%%%%%%%%%%%%%%%%%%%%%%%%%%%%%%

\newenvironment{acknowledgement}{\relax}{\relax}
\begin{acknowledgement}
\section*{Acknowledgements}
 We acknowledge fruitful discussions with Peter Braun-Munzinger, Johanna Stachel, Marek Gazdzicki, Mark Gorenstein, Klaus Reygers and Alice Ohlson. 
This work is part of and supported by the DFG Collaborative Research Center ``SFB 1225 (ISOQUANT)".
\end{acknowledgement}

% \appendix
% \section{Distributions of the identity variables}
% %%%%%%%%%%%%%%%%%%%%%%%%%%%%%%%%%%%%%%%%%%%%%%%%%%%%%%%%%%%%%%%%%%%%%%%%%%%%%%%%%%%%
% \begin{figure}[H]
%   \centering
%   \includegraphics[width=11cm]{\imagefolder/W_N_omega}
%   \caption{(Color online) The distribution of track variables $\omega$ (first column), event variables $W$ (second column) and the true 
%   multiplicity distributions $N$ (third column) of electrons, pions, kaons, protons and deuterons.}
%   \label{fig:WsNs}
% \end{figure}

%% References


\bibliographystyle{model1-num-names}
\bibliography{elsarticle-template-1-num.bib}






% \end{thebibliography}

\end{document}